\documentclass[12pt]{iopart}

\usepackage{bm}
\usepackage{epsfig}
\eqnobysec

\newcommand{\lwig}{\mbox{\;\raisebox{.3ex}
    {$<$}$\!\!\!\!\!$\raisebox{-.9ex}{$\sim$}\;}}
\newcommand{\gwig}{\mbox{\;\raisebox{.3ex}
    {$>$}$\!\!\!\!\!$\raisebox{-.9ex}{$\sim$}}\;}
\newcommand{\lambdabar}{{\hbox{$\lambda$\kern-1.ex\raise+0.45ex\hbox{--}}}}

\begin{document}

\begin{flushright}
{\large \tt DESY 05-079}
\end{flushright}

\title{Is it possible to tell the difference between fermionic and
bosonic hot dark matter?}

\author{Steen~Hannestad}
\address{Department of Physics and Astronomy \\
University of Aarhus, DK-8000 Aarhus C, Denmark}

\author{Andreas~Ringwald}
\address{Deutsches Elektronen-Synchrotron DESY, Hamburg,
Germany}

\author{Huitzu~Tu}
\address{Department of Physics and Astronomy \\
University of Aarhus, DK-8000 Aarhus C, Denmark}

\author{Yvonne~Y.~Y.~Wong}
\address{Deutsches Elektronen-Synchrotron DESY, Hamburg,
Germany}

\ead{\mailto{sth@phys.au.dk}, \mailto{andreas.ringwald@desy.de},
\mailto{huitzu@phys.au.dk}, \mailto{yvonne.wong@desy.de}}

\begin{abstract}
We study the difference between thermally produced fermionic and bosonic
hot dark matter in detail.  In the linear regime of
structure formation, their distinct free-streaming behaviours can
lead to pronounced differences in the
matter power spectrum.
While not detectable with current cosmological data,
such differences will be clearly observable with upcoming large scale
weak lensing surveys for particles as light as
$m_{\rm HDM} \sim 0.2 \ {\rm eV}$.
In the nonlinear regime,
bosonic hot dark matter is
not subject to the same phase space constraints that severely
limit the amount of fermionic hot dark matter infall
 into cold dark matter halos.  Consequently, the overdensities in fermionic and
bosonic hot dark matter of equal particle mass can differ by
 more than a factor of five in the central part of a halo.  However,
 this unique manifestation of quantum statistics may prove very difficult
 to detect unless the mass of the hot dark matter particle and its decoupling
 temperature fall within a very narrow window,
 $1 \lwig m_{\rm HDM}/{\rm eV} \lwig 4$ and $g_{*} \lwig 30$.  In this case,
hot dark matter infall
may have some observable consequences for the nonlinear power spectrum and hence
 the weak lensing convergence power spectrum
 at $\ell \sim 10^3 \to 10^4$ at the percent level.
\end{abstract}
\maketitle

\section{Introduction}

With the advent of precision measurements of the cosmic microwave
background (CMB), large scale structure (LSS) of galaxies, and
distant type Ia supernovae, a new paradigm of cosmology has been
established. In
this new standard model, the geometry is flat so that
$\Omega_{\rm total} = 1$, and the total
energy density is made up of
matter ($\Omega_m \sim
0.3$) [comprising of
baryons ($\Omega_b \sim
0.05$) and cold dark matter ($\Omega_{\rm CDM} \sim 0.25$)], and dark
energy ($\Omega_X \sim 0.7$). With only a few free parameters this
model provides an excellent fit to all current observations
\cite{Riess:1998cb,Perlmutter:1998np,Spergel:2003cb,bib:sdss1}.
In turn, this allows for constraints on other, nonstandard
cosmological parameters. One very interesting possibility
discussed widely in the literature is a subdominant
contribution to the total energy density in the form of
neutrino hot dark matter (HDM)
and hence limits on the neutrino mass.

In general, any fermionic dark matter species $\psi$
which decouples while relativistic obeys the relation
\begin{equation}
\label{eq:omegahdm}
\Omega_\psi h^2 = g \times \frac{10.75}{g_{*,\psi}} \times
\frac{m_\psi}{183 \,{\rm eV}},
\end{equation}
where $m_{\psi}$ is the $\psi$ particle's mass, $g$ its
number of internal degrees of freedom, and
$g_{*,\psi}$ the effective number of degrees of
freedom in the plasma at the time of $\psi$ decoupling.  By
entropy conservation, $g_{*,\psi}$  can
be related to the effective temperature of the species via
$g_{*,\psi}T_\psi^3=g_{*,\nu}  T_\nu^3$,
with the subscript $\nu$ referring  to the
standard model neutrinos.
Standard model neutrinos decouple in the early universe at a
temperature of order $2\to 3$ MeV, when $g_* = 10.75$. Thus,
just from
demanding  $\Omega_\nu \lwig \Omega_{\rm total} \simeq 1$,
one finds the well known
upper limit on the neutrino mass \cite{Gershtein:gg,Cowsik:gh},
$m_\nu \lwig 46/N_{\nu} \ {\rm eV}$,
assuming $N_{\nu}$ neutrino flavours with degenerate masses.

However, a much stronger limit on the neutrino mass
can be derived by noticing that the
thermal history of HDM is very different from that of
CDM.
By definition HDM becomes nonrelativistic only
at very late times.  At early times free-streaming of
the HDM particles
causes essentially all of their own perturbations to be erased on
scales below
the free-streaming length \cite{kolb},
\begin{equation}
\lambda_{\rm FS} \sim \frac{20~{\rm Mpc}}{\Omega_\psi h^2}
\left(\frac{T_\psi}{T_\nu}\right)^4 \left[1+\log \left(3.9
\frac{\Omega_\psi h^2}{\Omega_m h^2}
\left(\frac{T_\nu}{T_\psi}\right)^2 \right)\right]\,,
\label{eq:freestream}
\end{equation}leaving only
perturbations in the nonrelativistic matter (CDM and baryons).
On scales larger than $\lambda_{\rm FS}$, however,
HDM behaves like CDM.
Thus, the net result is a suppression of the overall level of
fluctuations on scales below $\lambda_{\rm FS}$.

In terms of the present matter power spectrum,
$P(k,\tau) \equiv |\delta|^2(k,\tau)$,
where $\delta (k,\tau)$ is the Fourier transform of the density
perturbations $\delta (x,\tau)$, early free-streaming
leads to a suppression of power at $k \gg 2 \pi/\lambda_{\rm FS}$
by roughly
\begin{equation}
\label{eq:deltap}
\frac{\Delta P(k,\tau)}{P(k,\tau)}
\simeq -8 \times
\frac{\Omega_\psi}{\Omega_m},
\end{equation}
where the factor eight is
derived from a numerical solution of the Boltzmann equation \cite{Hu:1997mj}.
Equation (\ref{eq:deltap}) applies only when $\Omega_\psi
\ll \Omega_m$; when $\Omega_\psi$ dominates, the spectrum
suppression becomes exponential as in the pure HDM
model.
Note that the free-streaming scale
becomes very small for very large values of
$g_{*}$ ($\sim 10^3$). In this case, the species should be
better known as
warm dark matter (see \cite{lesgourgues} for a recent
discussion of mass bounds on warm dark matter).

Of course, a suppression of small scale power in the present day
matter power spectrum could have its roots in a number of other
factors, such as a lower matter density, a higher relativistic
energy density, a primordial power spectrum with broken scale
invariance, etc..  Therefore,  in order to constrain the HDM
energy density and hence the mass of the HDM particle,
it is necessary to combine LSS with CMB measurements.
Currently available cosmological data constrain
the sum of light, standard model neutrino masses to
$\sum m_\nu \simeq 0.5\to 2$ eV, depending both on
the data sets used and on assumptions about other cosmological
parameters
\cite{Spergel:2003cb,bib:hannestad2003,Elgaroy:2004rc,Barger:2003vs,%
Crotty:2004gm,Seljak:2004xh,Fogli:2004as,Tegmark:2005cy,Hannestad:2005gj}.

While cosmological limits on HDM have mainly
been discussed for neutrinos, they apply equally well to bosons
and thermal relics with $g_{*}>10.75$
\cite{Hannestad:2003ye}. Many such hypothetical HDM
candidates exist, including the likes of axions
\cite{Moroi:1998qs,Hannestad:2005df}, majorons, gravitinos
\cite{Ibe:2005xc}, axinos \cite{Brandenburg:2004du}, etc..
An even more exotic scenario recently
discussed in the literature is the
possibility that neutrinos  violate Fermi--Dirac
spin statistics and behave like bosons
\cite{Dolgov:2005qi,Dolgov:2005mi}.
It is clearly worthwhile to study the
differences between these various HDM species arising
from their quantum statistics. As we shall see
later, there are fundamental differences in their clustering
properties which might be observable with future, high precision
probes.

Indeed, the purpose of this paper is to study the differences
between fermionic and bosonic HDM. From here on we
consider two generic HDM species: A single massive
Majorana fermion $\psi$ with $g=2$, and a single massive scalar $\phi$
with $g=1$.
Unless otherwise indicated, we shall assume, throughout the present work,
 a cosmological constant $\Lambda$,
three massless standard model neutrinos, and the following
cosmological parameters:
$\{\Omega_m,\Omega_{\Lambda},\Omega_b,h,\sigma_8\}=\{0.3,0.7,0.05,0.7,0.9\}$.

The paper is structured as follows.
In the next section we discuss structure formation in the linear
regime. Section \ref{sec:nonlinear}
 deals with nonlinear structure
formation, including hot dark matter infall into CDM halos.
We consider gravitational lensing as a probe of structure
formation in section \ref{sec:observation}, and section
\ref{sec:conc} contains our conclusions.

\section{Structure formation in the linear regime \label{sec:linear}}%

\begin{figure}
\hspace*{25mm}
\epsfxsize=10cm \epsfbox{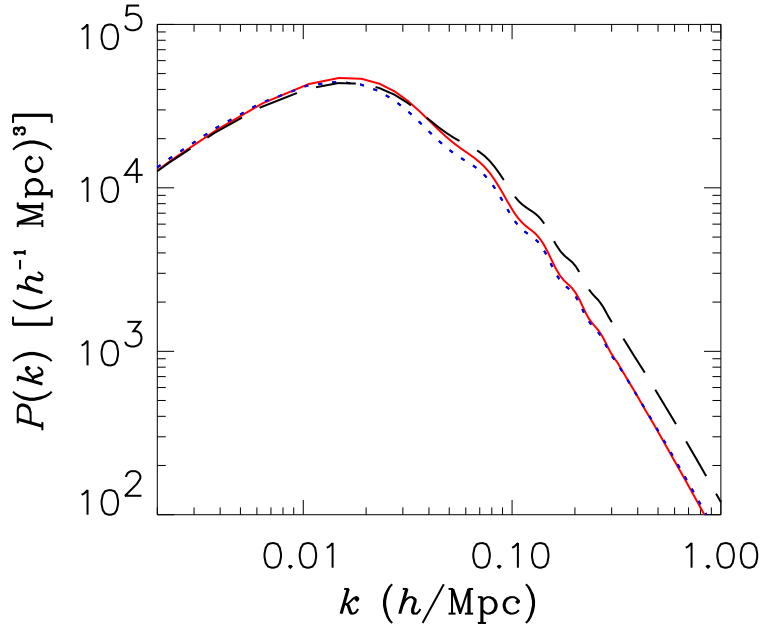}
\caption{\label{fig:pspec}Linear power spectra for two different
$\Lambda$HCDM models. The blue (dotted) line shows a model with
three massless neutrinos and one massive Majorana fermion,
contributing $\Omega_\psi = 0.02$. The red (solid) line shows the
same, but with a massive scalar instead. The black (dashed)
line is the standard $\Lambda$CDM model with no HDM.
Note that these spectra have been normalised to have the same
amplitude on large scales.}
\end{figure}

A thermal relic that decouples from the cosmic plasma while
relativistic assumes
either the relativistic Fermi--Dirac (FD, ``$+$'') or Bose--Einstein
(BE, ``$-$'')
distribution,
\begin{equation}
f(p) = \frac{1}{\exp(p/T) \pm 1}.
\end{equation}
If the particle is stable and no
heavier particle decays into it, its
phase space distribution is preserved with the expansion
of the universe.
Today, a single massive $g=2$ Majorana fermion $\psi$ and a $g=1$ scalar
$\phi$ are expected to contribute to the total energy density at fractions
of
\begin{eqnarray}
\Omega_\psi h^2 &=& 2 \times \frac{10.75}{g_{*,\psi}} \times
\frac{m_\psi}{183 \,{\rm eV}}, \\
\Omega_\phi h^2 &=& \frac{4}{3} \times 1 \times \frac{10.75}{g_{*,\phi}}
\times \frac{m_\phi}{183 \,{\rm eV}}.
\end{eqnarray}
Already, we can see that for any given $\Omega_{\rm HDM}$, the mass of the
scalar is necessarily a factor $3/2$ higher than that of the fermion.

With regard to structure formation, the amount of
suppression in the matter power spectrum due to HDM
is essentially dependent only
on the ratio $\Omega_{\rm HDM}/\Omega_m$, i.e., it depends only on
the lack of dark
matter. Therefore, on both very small and very large scales,
fermion and scalar models with identical $\Omega_{\rm HDM}$ are
virtually indistinguishable to the naked eye. However, for scales close to the
free-streaming length (and therefore also to the scale of
matter--radiation equality for sub-eV hot dark matter), there is a
pronounced difference. The reason is that for the same $\Omega_{\rm HDM}$, the
more massive scalar particle exhibits less free-streaming.  This in turn allows
for an increased fluctuation amplitude around the free-streaming scale.
In Figure \ref{fig:pspec} we show the power spectra for two models with
identical $\Omega_{\rm HDM} = 0.02$ and $g_* = 10.75$
(corresponding to $m_\psi = 0.9$ eV for the fermion model and
$m_\phi = 1.3$ eV for the scalar model). As can be seen, the
scalar model has more power at $k \sim 0.01 \to 0.1 \ h \ {\rm Mpc}$. For
comparison we also show the standard $\Lambda$CDM model, which has
the same parameters, except that $\Omega_{\rm HDM} = 0$.

\section{Nonlinear structure formation \label{sec:nonlinear}} 

At late times ($z \lwig 100$), fluctuations in the mass density field
can become larger than unity.  Linear perturbation theory breaks down,
and structure formation enters a nonlinear phase with the
collapse of overdense regions into gravitationally bound objects.
In hierarchical CDM cosmologies, nonlinear gravitational
collapse begins at small scales, and the nature of the CDM particle
determines the size of the first objects.
Subsequent mergers of these small systems give rise to  larger
structures---from the halos of dwarf galaxies to the filaments and
the voids.
In terms of Fourier decomposition,
the transition from the linear to the nonlinear regime in
the density field occurs roughly
at $k \sim 0.2 \ h \ {\rm Mpc}^{-1}$ today.

\begin{figure}
\hspace*{5mm}
\epsfxsize=14cm
\epsfbox{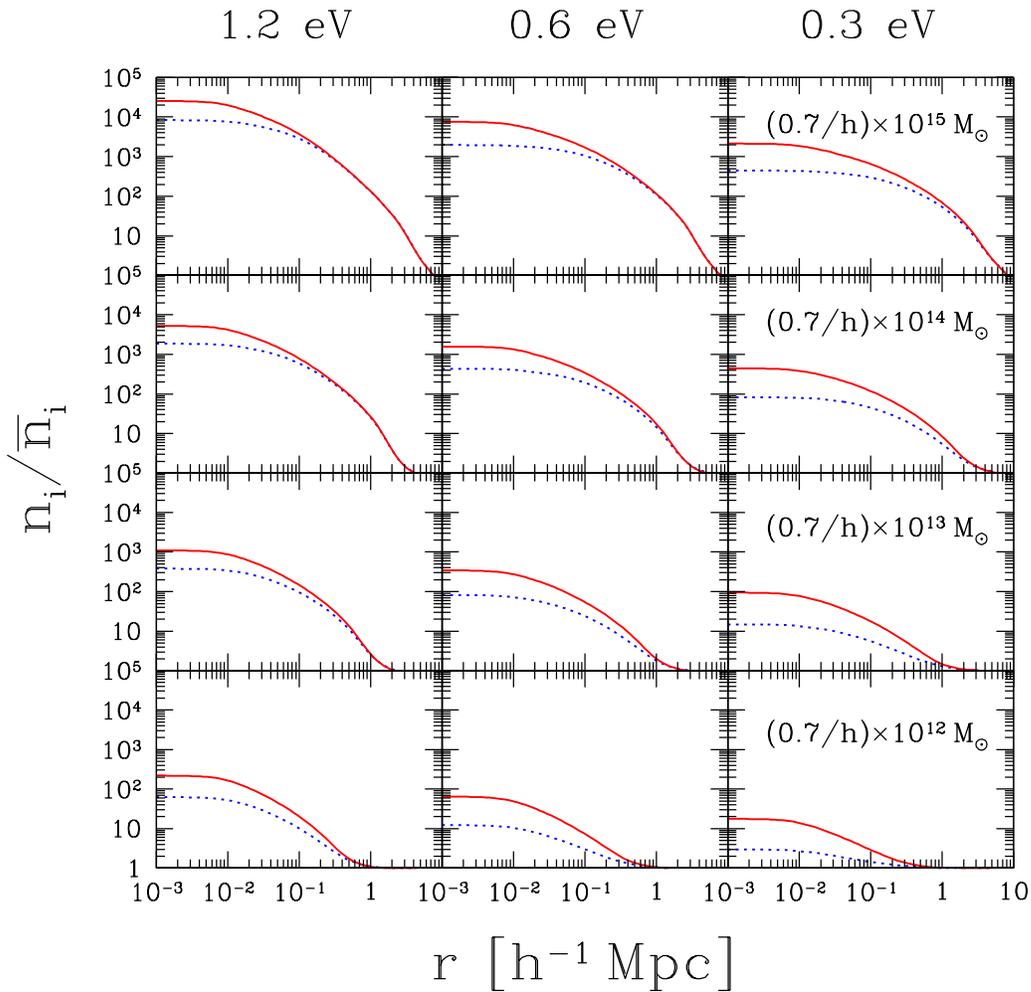}
\caption{Number densities of HDM expected for $g_*=10.75$ and the
indicated halo and HDM masses at $z=0$, normalised to
the cosmological average $\bar{n}_i$. Red (solid) lines represent
bosons, while blue (dotted) lines denote
fermions.
These densities are obtained from
solving the Vlasov equation
using the method of \cite{Ringwald:2004np},
assuming
CDM halos with the Navarro--Frenk--White density
profile
[equation (\ref{eq:nfw})].
\label{fig:overdensities}}
\end{figure}

The effect of HDM on nonlinear clustering and its statistics
is twofold.
Firstly, free-streaming of HDM in the early (linear) stages of structure 
formation suppresses density fluctuations on scales below $\lambda_{\rm FS}$.
This directly limits the number of small scale structures that
can be formed subsequently in the  nonlinear phase.

\begin{figure}
\epsfxsize=7.5cm
\epsfbox{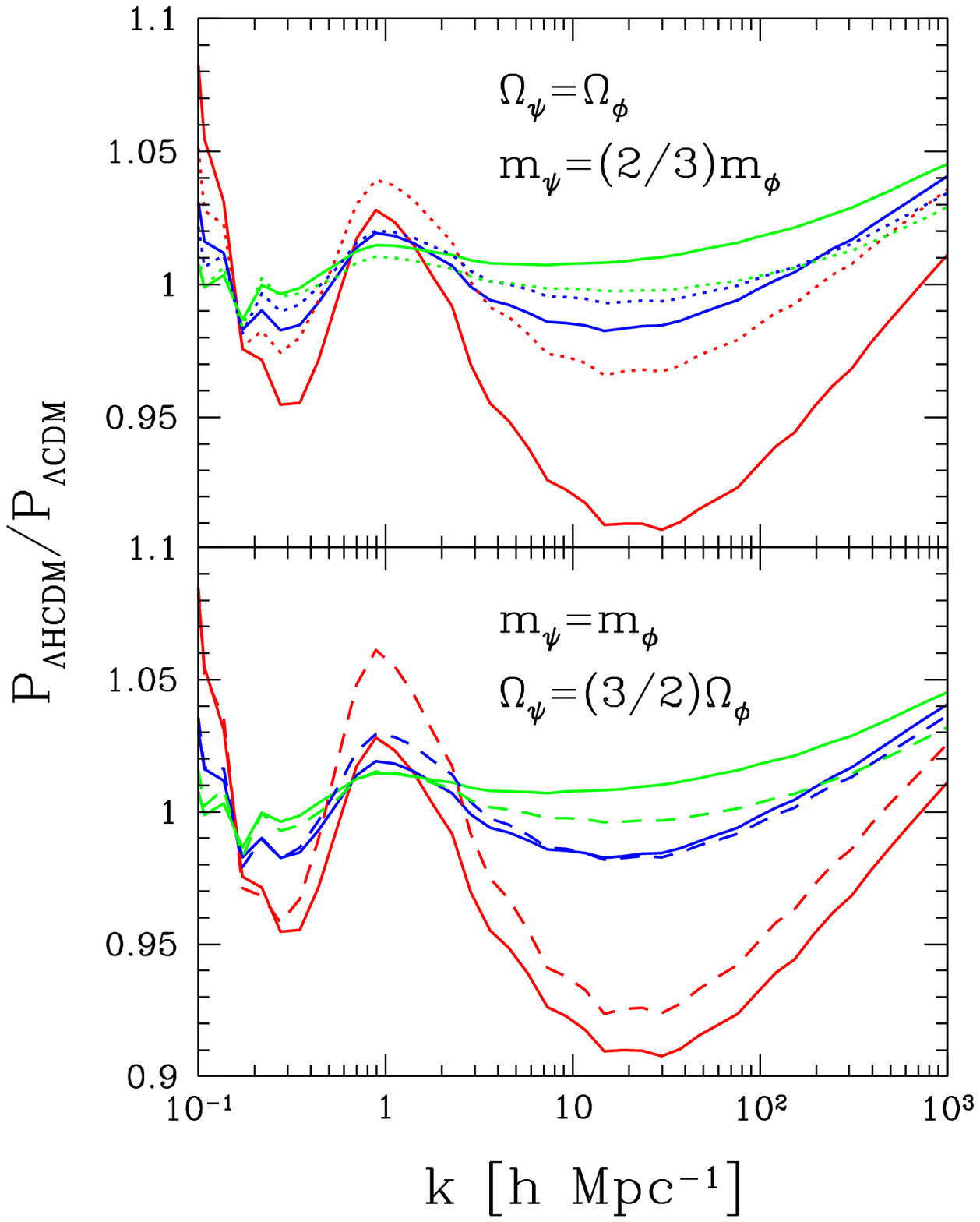}
\epsfxsize=7.5cm
\epsfbox{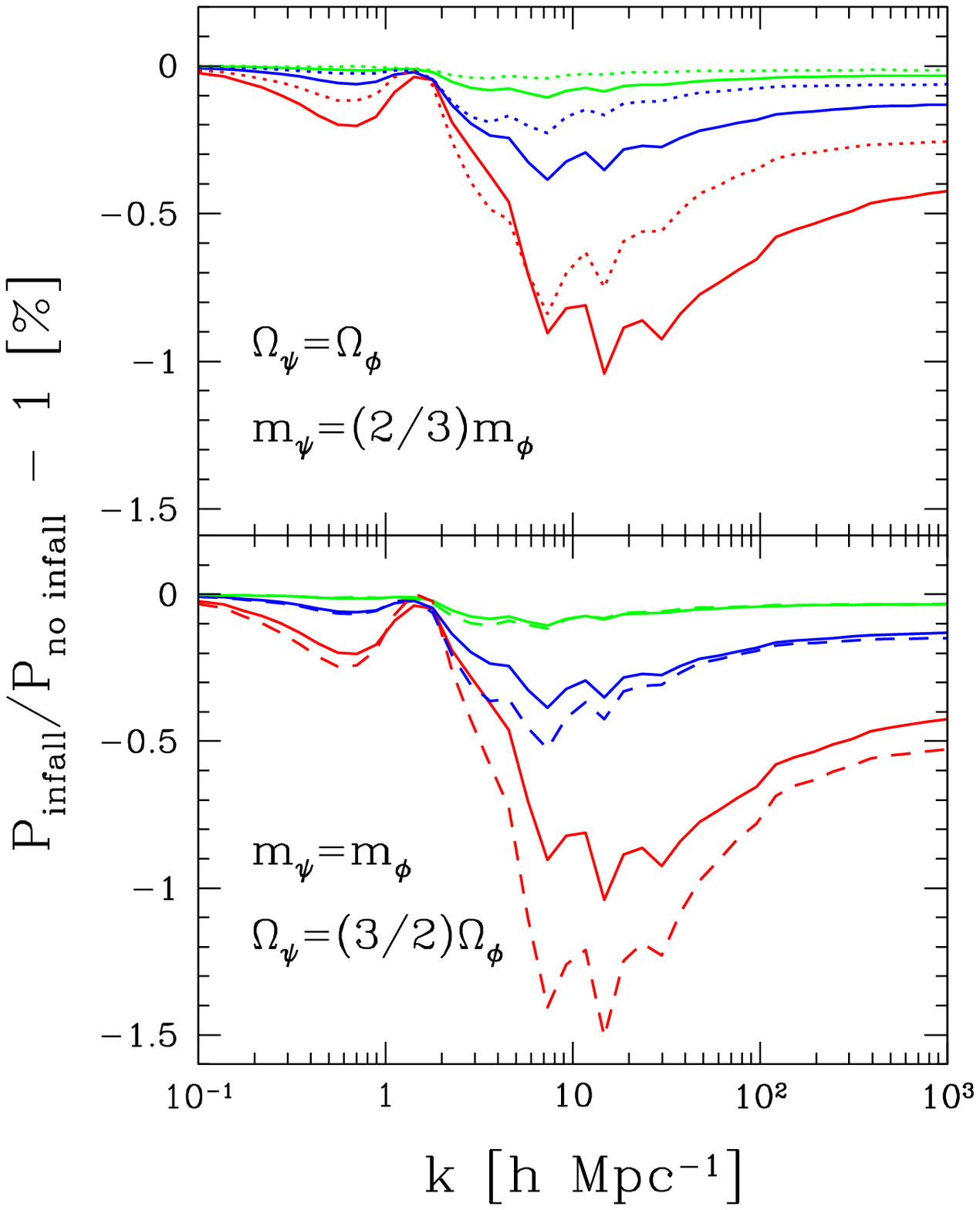}
\caption{{\it Left}: Nonlinear power spectra  at $z=0$ for
$g_*=10.75$ and various HDM masses, relative to the standard
$\Lambda$CDM power spectrum.  Solid lines
denote bosons of masses (top to bottom)
$\{0.3, 0.6, 1.2\} \ {\rm eV}$, dotted lines are for
$\{0.2, 0.4, 0.8\}\ {\rm eV}$ fermions, while
dashed lines represent
$\{0.3, 0.6, 1.2\} \ {\rm eV}$ fermions.  These masses are chosen
such that they satisfy either
$\Omega_{\psi}=\Omega_{\phi}$ or $m_{\psi}=m_{\phi}$.
Note that all spectra here have been normalised to a
fixed value of $\sigma_8$.
{\it Right}: Relative contributions from HDM infall for the same HDM
scenarios.
 \label{fig:infall}}
\end{figure}

Secondly, at late times, secondary infall of HDM into existing CDM halos
becomes possible when the former's mean velocity,
\begin{equation}
\label{eq:fdspeed}
\langle v \rangle \simeq  f \times 23 \ (1+z) 
\left(\frac{10.75}{g_{*}}\right)^{1/3}\!\! \left(\frac{\rm eV}{m_{\rm HDM}}
\right)\ {\rm km} \
{\rm s}^{-1}, \quad f=\left\{\begin{array}{ll} 7 & {\rm FD} \\
6 & {\rm BE} \end{array} \right.
\end{equation}
drops below the velocity dispersion of the astrophysical system concerned.
A typical galaxy cluster ($M \sim 10^{14} \to 10^{15} \ M_{\odot}$)
has a velocity dispersion of about
$1000 \ {\rm km \ s}^{-1}$ today; a typical galaxy ($M \sim 10^{12} \ 
M_{\odot}$), about
$200 \ {\rm km \ s}^{-1}$.  Thus, for $m_{\rm HDM} \sim 1 \ {\rm eV}$,
a good fraction of HDM particles can be expected to reside
presently in
halos across a wide mass range.
Figure \ref{fig:overdensities} shows a sample of
HDM overdensities
expected for a scalar and a Majorana fermion
for a range of halo masses.
These are obtained from solving the Vlasov equation (\ref{eq:vlasov}), assuming
CDM halo density profiles of the Navarro--Frenk--White form
[equation (\ref{eq:nfw})]
\cite{Navarro:1995iw,bib:nfw}.
A brief description of this calculation can be found in
\ref{clustering}.  For more details, see reference \cite{Ringwald:2004np}.

Observe in Figure \ref{fig:overdensities}
that at the sub-Mpc scales, bosons cluster considerably more
efficiently than do fermions.  This is
because, at any given  temperature, the unperturbed phase space
distribution for bosons is more skewed towards the low momenta, making the
particles
more susceptible to capture by the CDM halos.
Furthermore, the final HDM phase space distribution in a halo
is, in general,  subject to constraints imposed by Liouville's theorem,
and hence must not exceed the maximum of the
initial unperturbed
distribution function, $f_{\rm final} \lwig f_{\rm initial}^{\rm max}$
\cite{tremaine1}.
For fermions with no chemical potential,
the maximum $f_{\rm initial}^{\rm max}=1/2$ occurs at $p=0$.
Using this line of argument, Tremaine and Gunn have famously derived
a bound on the maximum possible neutrino density in a halo
\cite{Tremaine:1979we,Kull:1996nx}.
Bosons, however, are technically not subject to the same constraint,
since no finite maximum exists in their initial
phase space distribution.  Thus, correspondingly in
Figure \ref{fig:overdensities},
the overdensity of bosons is seen to be  much higher than that
of fermions with the same mass.

Note, however, that the Tremaine--Gunn bound is still applicable to bosons in
a statistical sense because only a very small fraction of all
thermally produced bosons reside in low momentum states
\cite{Madsen:1990pe,Madsen:1991mz}. (One notable exception is axion
cold dark matter which consists of low mass bosons in a
condensate.)

\begin{figure}
\epsfxsize=15.7cm
\epsfbox{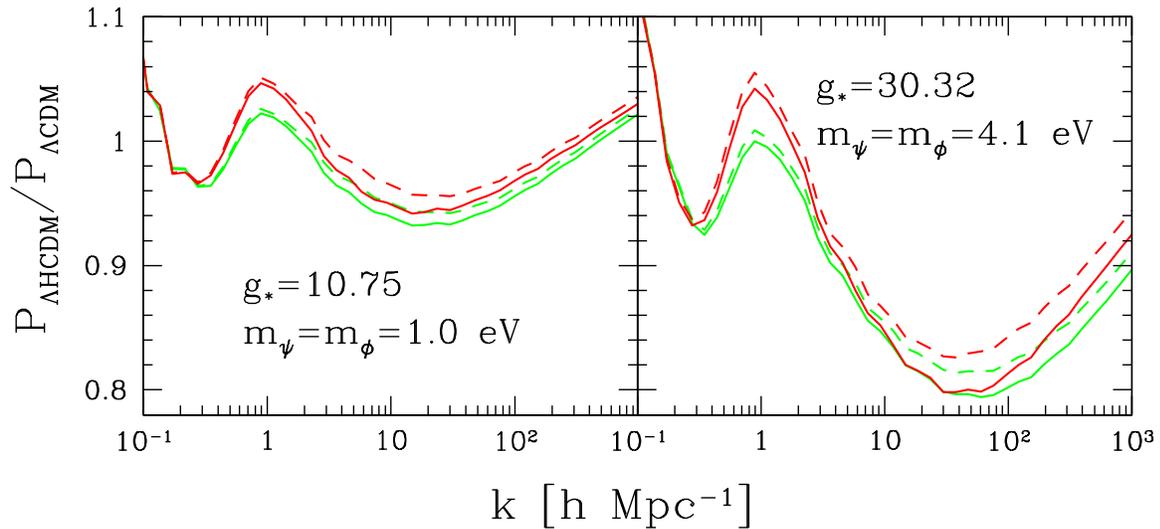}
\caption{Nonlinear power spectra at $z=0$ for two different values of $g_{*}$,
relative to
the standard $\Lambda$CDM power spectrum.  Red (dark) lines denote fermions,
while green (light) lines represent bosons.  Solid lines show the power spectra
with HDM infall, dashed lines show them without.
\label{fig:gstar_fd_be}}
\end{figure}

\subsection{The nonlinear power spectrum}

\begin{figure}
\hspace*{23mm}
\epsfxsize=11cm
\epsfbox{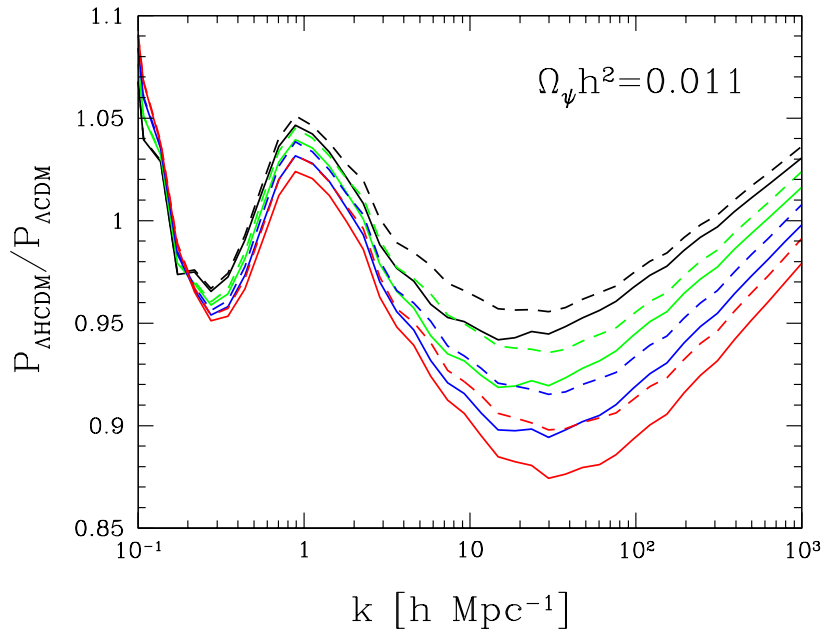}
\caption{Nonlinear power spectra at $z=0$ for fermions with various values of
$g_{*}$ and $m_{\psi}$ but the same $\Omega_{\psi}h^2=0.011$,
relative to the standard $\Lambda$CDM power spectrum.
Solid lines show the power spectra with HDM infall,
dashed lines without. From top to bottom: $m_{\psi}=\{1, 1.41, 2.00, 2.82\} \ 
{\rm eV}$, corresponding to $g_{*}=\{10.75,15.16,21.5,30.32\}$.
\label{fig:gstar_fd}}
\end{figure}

In Figures \ref{fig:infall} to \ref{fig:gstar_fd}, we show predictions
for the nonlinear power spectrum at $z=0$ for a variety of scenarios involving
bosonic and fermionic
HDM.  These spectra are calculated from a phenomenological
approach known as the halo model
\cite{Seljak:2000gq,Peacock:2000qk,Ma:2000ik,Cooray:2002di},
extended to include  contributions from HDM infall
\cite{Abazajian:2004zh}.  While the halo model will not
serve for high precision cosmology, it suffices to illustrate
the typical scales of the various effects due to HDM.
See \ref{halomodel} for more details.

As is evident in the Figures,
early HDM free-streaming is generally
the dominant influence on the
shape of the nonlinear power spectrum.
Ignoring contributions from late-time infall, one can see
that early free-streaming alone already distinguishes
between fermions and bosons,
and between cases with different $g_{*}$'s
at the percent level.
Late-time HDM infall causes additional losses of power at
$k \gwig 1 \ h \ {\rm Mpc}^{-1}$,
because HDM clustering
tends to ``smooth out'' the overall matter
distribution within the individual halo.
For the $g_*=10.75$ cases (Figure \ref{fig:infall}),
losses from infall are
typically no more than one percent,
and
occur at scales corresponding to
roughly  ten times the {\it current} free-streaming wavenumber
of the particle $k_{\rm FS}$,
\begin{equation}
\label{eq:currentkfs}
k_{\rm FS} \simeq f \times 0.11 \ \sqrt{\frac{\Omega_m}{1+z}}
\left(\frac{g_{*}}{10.75}\right)^{1/3}
\left(\frac{m_{\rm HDM}}{\rm eV} \right) \ h \ {\rm Mpc}^{-1}.
\end{equation}
The analysis of \cite{Abazajian:2004zh} also finds
a reduction of power due to infall
at the $0.1 \to 1 \ \%$ level, albeit at a scale of
$k\sim 0.5 \ h \ {\rm Mpc}^{-1}$.
This discrepancy
could be due to the linear method used in their clustering
calculations.

Infall increases in importance for cases with  $g_* >10.75$.  This is
because these particles are colder,
and for the same $\Omega_{\rm HDM}$,
are heavier; both attributes allow
them to cluster more efficiently.  Indeed, as shown in Figures 
\ref{fig:gstar_fd_be} and \ref{fig:gstar_fd}, HDM infall in any one
scenario can
alter the resulting power spectrum to the extent that it mimics another scenario,
and is therefore potentially important for high precision cosmology.
Examples of thermal relics with large $g_*$'s include very light gravitinos
($g_*\sim 100$, corresponding to a decoupling temperature of a few GeV to a few 
TeV),
and thermally produced axions ($10.75 \lwig g_* \lwig 80$, corresponding to
decoupling during the QCD phase transition).

\section{Observational probes \label{sec:observation}}

\subsection{Weak gravitational lensing}

Weak gravitational lensing of distant galaxies by intervening
large
scale structure provides a unique method to directly
map the matter distribution  in the universe
(e.g., \cite{Bartelmann:1999yn}).
Perturbations in the gravitational potential along the line of sight
induce
distortions in the images of distant galaxies.
The images may be magnified (convergence) and/or stretched (shear).
For weak lensing of a large ensemble of sources and lenses,
it can be shown that the convergence and the shear
have identical statistical
properties.  Henceforth, we shall consider the power
spectrum of the convergence as representative
of the lensing features.

The convergence is essentially an integral over all the
deflectors between us and the source galaxies, weighted by
the source galaxy distribution.
Its power spectrum at multipole $\ell$
can be related to the
matter power spectrum $P(k,z)$
via \cite{Bartelmann:1999yn,Kaiser:1991qi,Kaiser:1996tp}
\begin{equation}
C_{\ell} = \frac{9}{16}  \ H_0^4 \ \Omega_m^2 \int^{\chi_h}_0  d \chi
\left[\frac{g(\chi)}{a \chi} \right]^2 P \left( \frac{\ell}{\chi},z \right)\, ,
\end{equation}
in a flat universe. Here, $\chi  = \int^z_0 \  d z/H(z)$ is the comoving radial 
distance,
and $\chi_h$ is the distance to the horizon.
The weak lensing weighting function,
\begin{equation}
\label{eq:weight}
g(\chi) = \chi \int^{\chi_h}_{\chi} d \chi' \ n(\chi') \
\frac{\chi'-\chi}{\chi'}\, ,
\end{equation}
encapsulates information about the source galaxy redshift distribution
$n(\chi)$.
For our purposes, $n(\chi)$ may be taken to be
$n(z) \propto z^{\alpha} \exp[-(z/z_0)^{\beta}]$, where the
parameters $z_0$, $\alpha$ and $\beta$ are estimated from existing
deep redshift surveys for the weak lensing survey at hand
\cite{Massey:2003xd,Refregier:2003xe}.  The prefactor is determined
by the normalisation condition $\int d z \ n(z) = 1$.
In the following, we shall
take $z_0=1$, $\alpha=2$ and $\beta=2$, typical of
most proposed large scale lensing surveys \cite{Refregier:2003xe}.
With these parameters, the weak lensing weighting function
(\ref{eq:weight}) peaks at $z \sim 0.4 \to 0.5$.

\begin{figure}
\hspace*{23mm}
\epsfxsize=11cm
\epsfbox{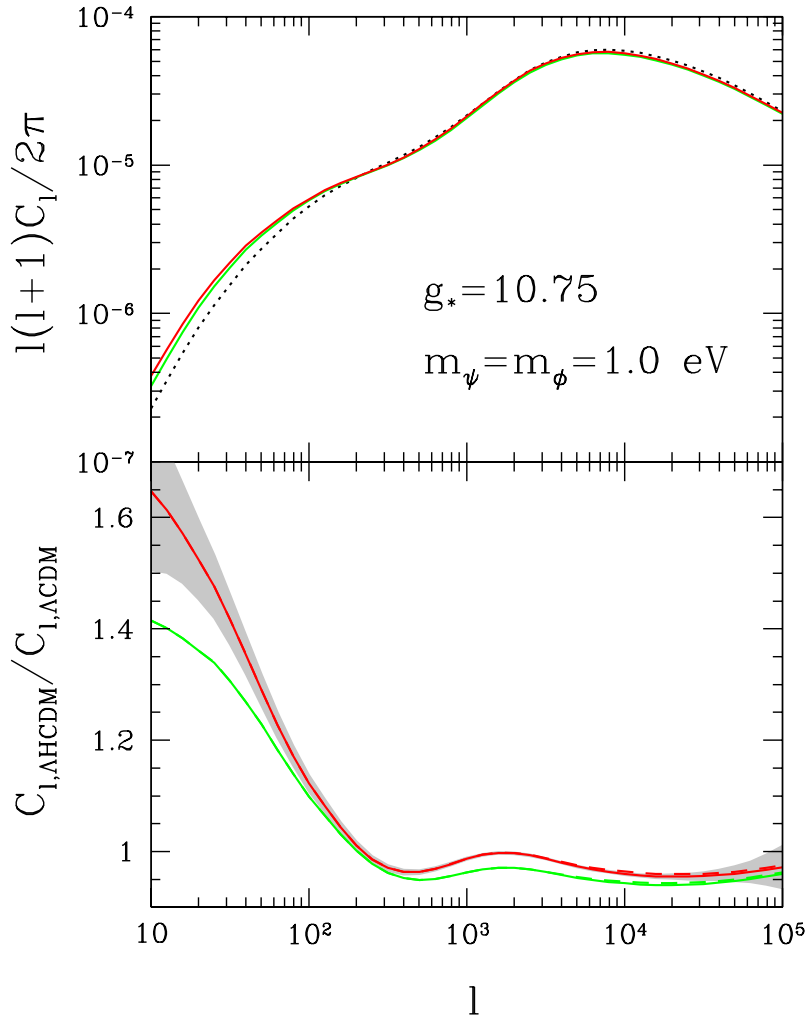}
\caption{{\it Top}: Weak lensing convergence power spectra for the models shown
in the left panel of Figure \ref{fig:gstar_fd_be}.  The
red (dark solid) line denotes fermions, the green (light solid) line
represents bosons,
while the black (dotted) line corresponds to the standard $\Lambda$CDM model.
{\it Bottom}: The same two convergence spectra normalised to the standard $\Lambda$CDM
convergence spectrum. Solid lines show the spectra with HDM infall,
dashed lines show them without.
The grey shaded area represents errors expected for a LSST-type survey,
 smeared over bands of width $\ell/4$.
Figure \ref{fig:lensed_2_zoom} left shows the same plot at $\ell = 10^2 \to 10^5$
for greater clarity.
\label{fig:lensed_2}}
\end{figure}

In any lensing survey, the ability
to measure $C_{\ell}$ is constrained on large scales by
a sample variance due to finite sky coverage, and on small scales by
the finite number of available galaxies.
Neglecting non-Gaussian corrections, the statistical error
in $C_{\ell}$ is estimated to be \cite{Kaiser:1991qi,Kaiser:1996tp}
\begin{equation}
\Delta C_{\ell} = \sqrt{\frac{2}{(2 \ell +1) f_{\rm sky}}}
\left(C_{\ell} +\frac{\gamma_{\rm rms}^2}{n_{\rm gal}} \right),
\end{equation}
where $f_{\rm sky}$ is the fraction of the sky covered by the survey,
$n_{\rm gal}$ is the surface density of galaxies, and $\gamma_{\rm rms}$
is the rms shear per galaxy (from noise and intrinsic ellipticity).
These parameters are generally survey dependent.
Proposed wide-field surveys
such as the
Supernova/Acceleration Probe
(SNAP) \cite{bib:snap} are expected to cover
some $1000 \ {\rm deg}^2$ ($f_{\rm sky} \sim 0.03$) of the sky;
the coverage of the more ambitious Large Synoptic Survey
Telescope (LSST) \cite{bib:lsst} will likely be another tenfold.
In the ensuing analysis, we shall adopt
$\gamma_{\rm rms}=0.15$, $n_{\rm gal}=50 \ {\rm arcmin}^{-2}$,
and $f_{\rm sky}=0.5$, parameter values corresponding to an optimal
 LSST-type survey.  After smearing over bands of width $\ell/4$, we
 have approximately
\begin{equation}
\Delta C_l \simeq 0.004\, \frac{1000}{\ell} \sqrt{\frac{0.25}{f_{\rm sky}}}
\left(C_{\ell}+\frac{\gamma^2_{\rm rms}}{n_{\rm gal}} \right)\,
\end{equation}
as the effective error  \cite{Zhan:2004wq}.

\begin{figure}
\epsfxsize=7.5cm
\epsfbox{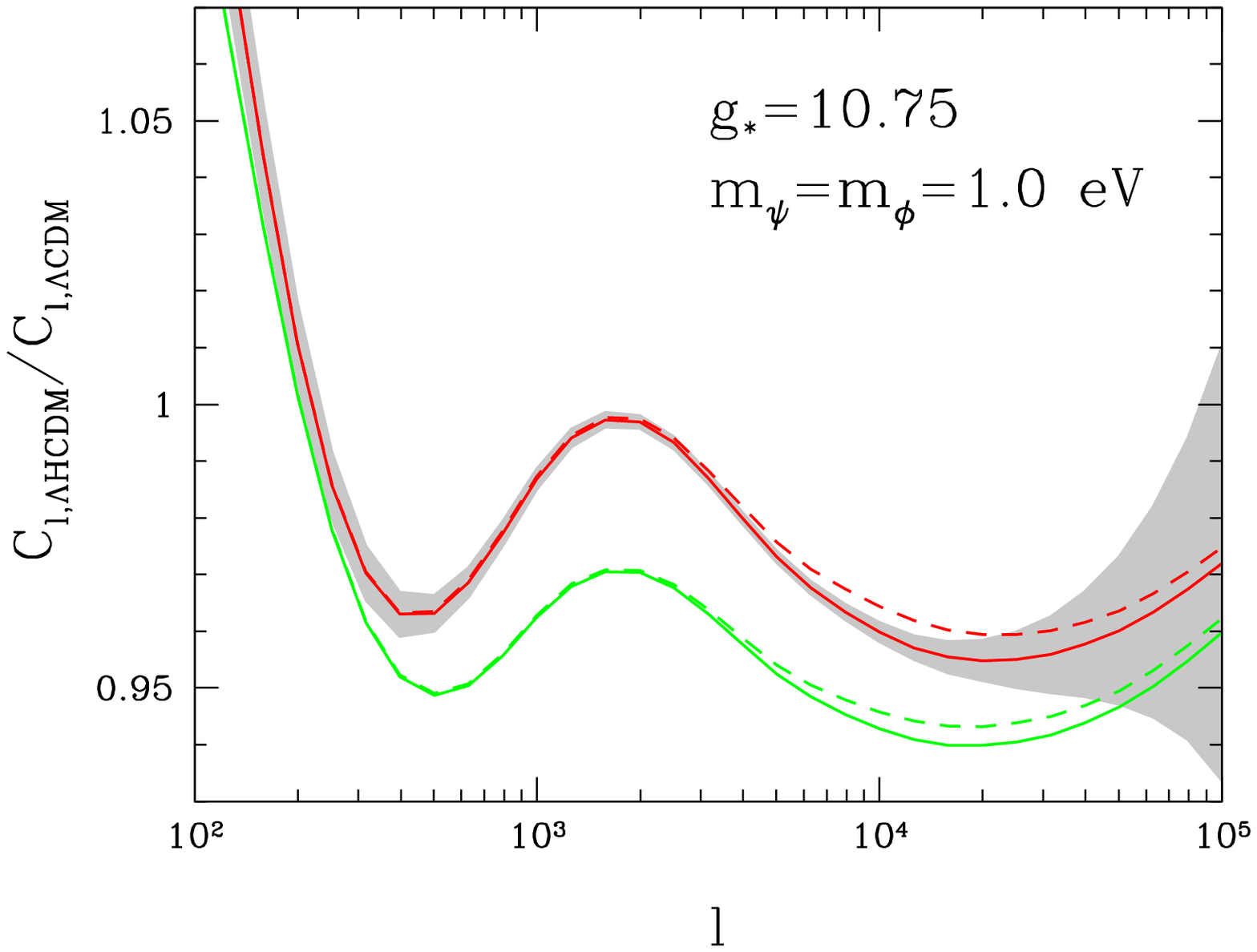}
\epsfxsize=7.5cm
\epsfbox{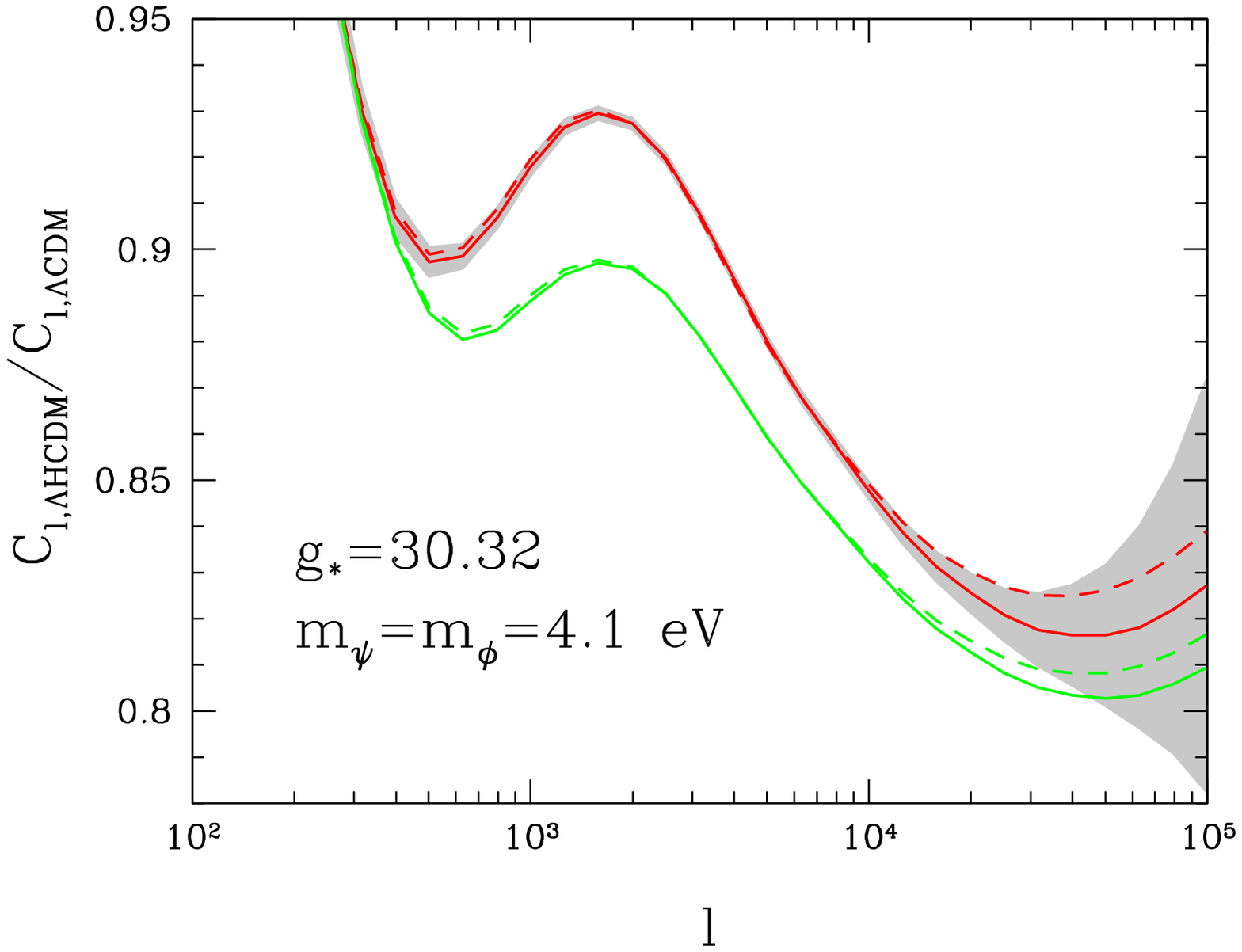}
\caption{Weak lensing convergence power spectra for the models shown
in Figure \ref{fig:gstar_fd_be}, relative to the standard
$\Lambda$CDM convergence spectrum.
Red (dark) lines denote fermions, while green (light) lines
represent bosons. Solid lines show the spectra with HDM infall,
dashed lines show them without. \label{fig:lensed_2_zoom}}
\end{figure}

\begin{figure}
\epsfxsize=7.5cm
\epsfbox{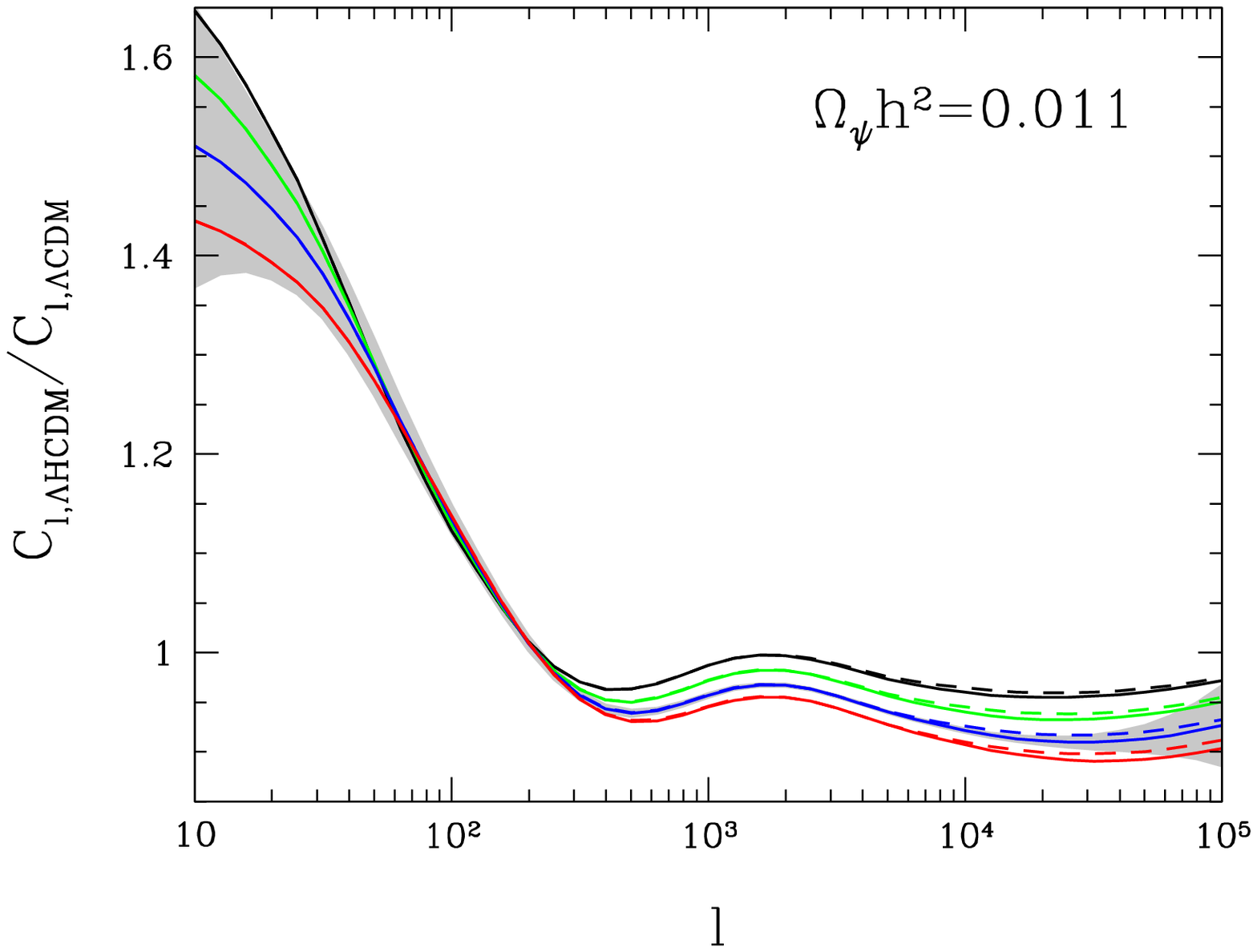}
\epsfxsize=7.5cm
\epsfbox{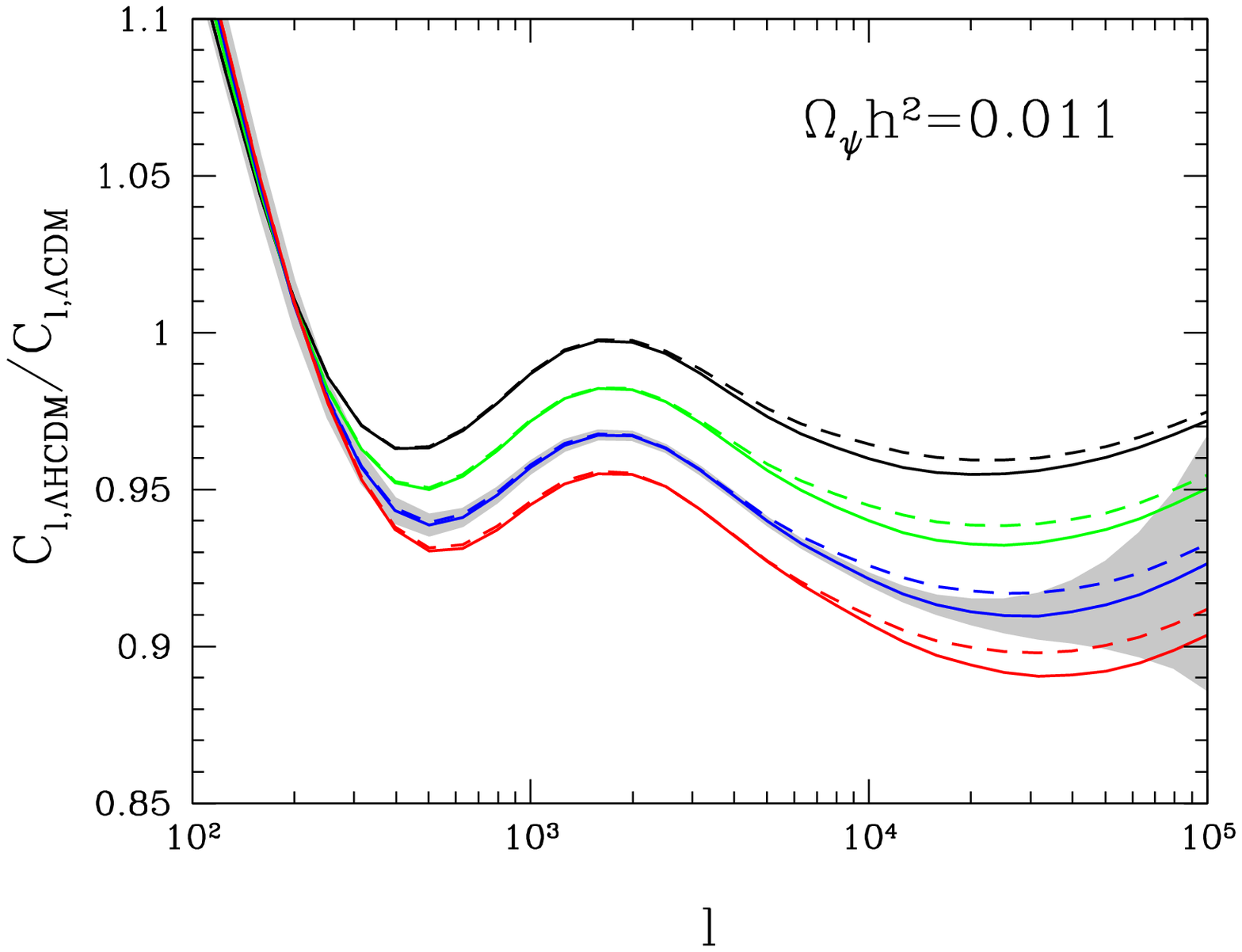}
\caption{{\it Left}: Weak lensing convergence power spectra for the
$\Omega_{\psi}h^2=0.011$ models shown
in Figure \ref{fig:gstar_fd}, relative to the standard $\Lambda$CDM
convergence spectrum. Solid lines show the spectra with HDM infall,
dashed lines without. The masses of the HDM particles are (top to bottom)
$\{1, 1.41, 2.00, 2.82\} \ {\rm eV}$,
corresponding to $g_{*}=\{10.75,15.16,21.5,30.32\}$.
{\it Right}: Same as the left panel, but zoomed in on $\ell = 10^2 \to 10^5$.
\label{fig:lensed}}
\end{figure}

\begin{figure}
\epsfxsize=7.5cm
\epsfbox{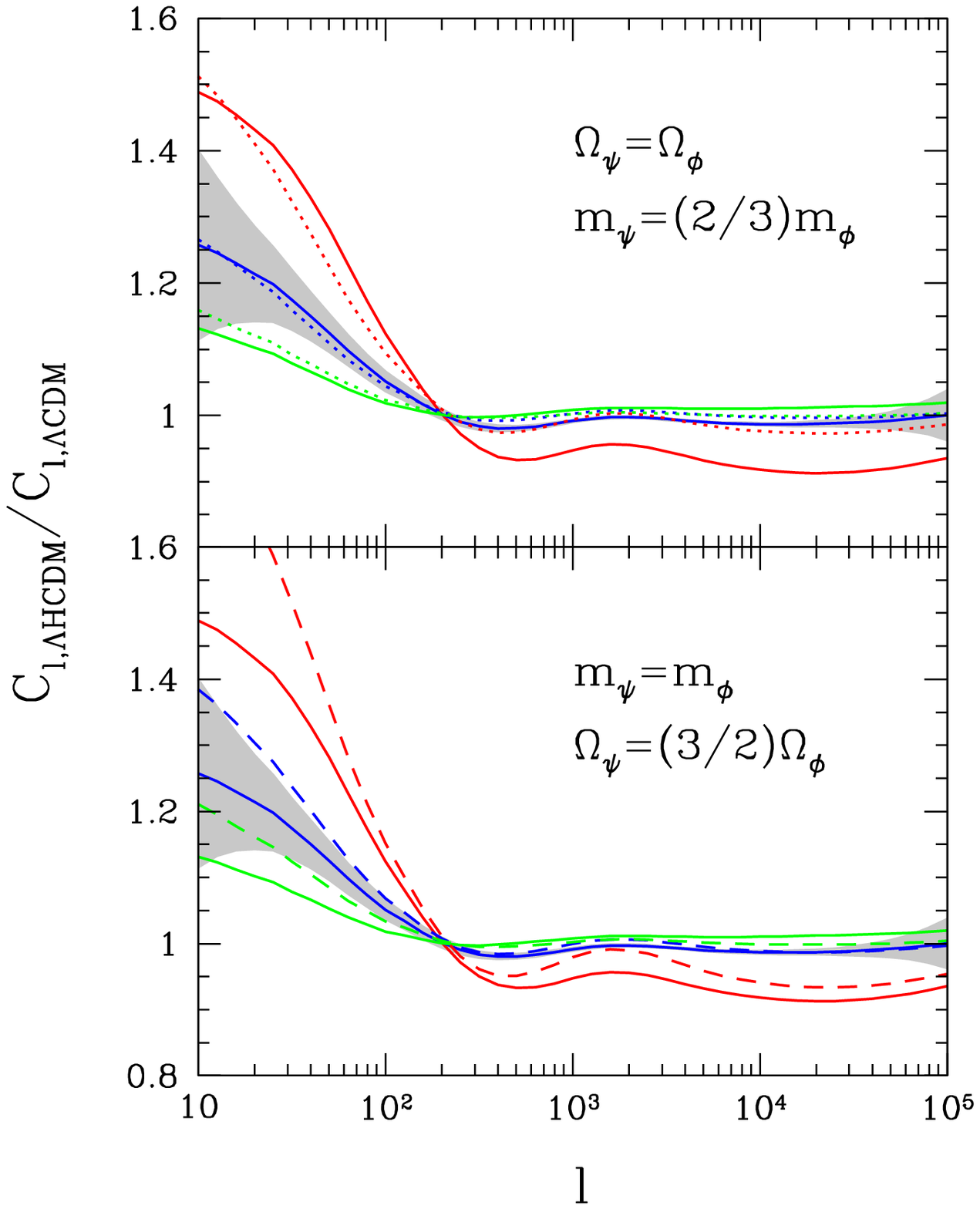}
\epsfxsize=7.5cm
\epsfbox{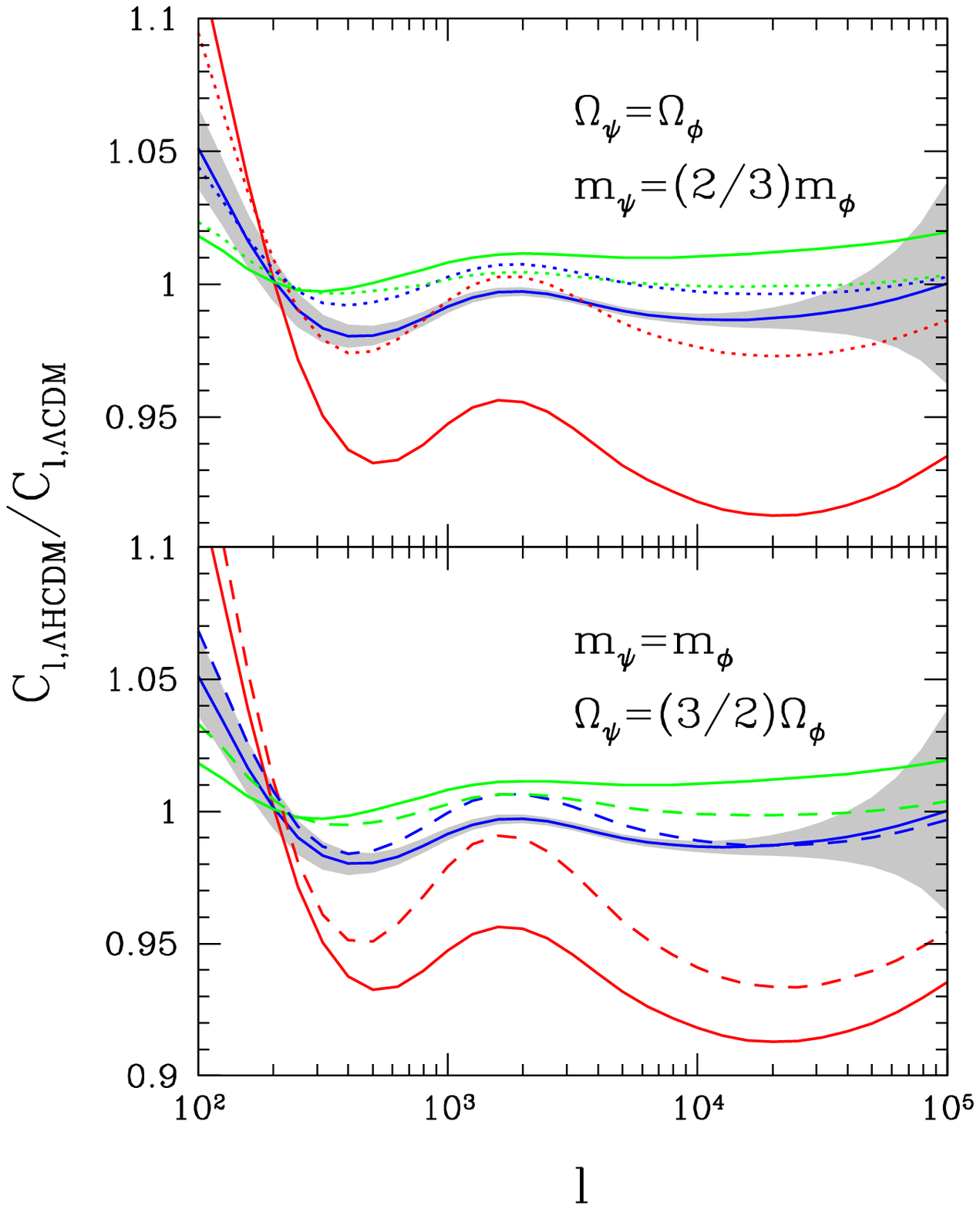}
\caption{{\it Left}: Weak lensing convergence power spectra for
the $g_*=10.75$ models shown in Figure \ref{fig:infall},
relative to the standard
$\Lambda$CDM convergence spectrum.  Solid lines
denote bosons of masses (top to bottom)
$\{0.3, 0.6, 1.2\} \ {\rm eV}$, dotted lines are for
$\{0.2, 0.4, 0.8\}\ {\rm eV}$ fermions, while
dashed lines correspond to
$\{0.3, 0.6, 1.2\} \ {\rm eV}$ fermions.
{\it Right}: Same as the left panel, but zoomed in on $\ell = 10^2 \to 10^5$.
 \label{fig:lensed_3}}
\end{figure}

Figures \ref{fig:lensed_2} to \ref{fig:lensed_3} show the expected weak lensing
convergence power spectra for  the various HDM scenarios considered
in the previous section.  The general shapes of the convergence
spectra are similar to those of their corresponding matter
power spectra shifted according to
$k \approx \ell / 1000$, and reflect predominantly the
early free-streaming behaviours of the different HDM species.
Contributions from infall, however, are considerably
reduced (by a factor of $2 \to 3$) relative to the
$z=0$ matter power spectrum.  This is because weak lensing is
primarily sensitive to the $z \sim 0.4 \to 0.5$ universe,
where the amount of HDM infall is correspondingly lower.

In Figures \ref{fig:lensed_2_zoom} and \ref{fig:lensed}, we see that
the contributions from infall are typically at the
sub-percent  level at $\ell \sim 10^3 \to 10^4$.  As with the
nonlinear power spectrum, the contributions increase with
$g_*$ and $m_{\rm HDM}$.  However, because the scale at which
HDM infall is manifest depends on
the {\it current} free-streaming scale of the HDM particle
[cf.\ equation (\ref{eq:currentkfs})],
the signatures of infall are also shifted accordingly
to larger values of $\ell$, where the statistical errors are
larger.  The net result is that there may only be a very narrow window
in $m_{\rm HDM}$ and $g_*$ in which
HDM infall has some observable consequences in weak lensing, even for
 an optimal LSST-type lensing survey.
 Judging from Figures \ref{fig:lensed_2_zoom} and \ref{fig:lensed}, we estimate
 this window to be $1 \lwig m_{\rm HDM}/{\rm eV} \lwig 4$ and $g_* \lwig 30$.

Nevertheless, fermionic and bosonic HDM can still be distinguished by their
 early free-streaming behaviours.  In Figure \ref{fig:lensed_3},
 we see that this is possible for an LSST-type lensing survey
  for particles as light as $m_{\rm HDM} \sim 0.2 \ {\rm eV}$.

\subsection{Strong gravitational lensing}

Looking at Figure \ref{fig:overdensities}, it is clear that the main
difference between fermionic and bosonic HDM is in the
central regions of halos. The matter distribution in the central
parts of dark matter halos can in principle be probed by
strong gravitational lensing, either by very precise measurements of a single
lensing system \cite{Maller:1999de,Trott:2002cr,Trott:2003mn}, or
by measuring a very large sample of systems (for instance using a
supernova survey \cite{Mortsell:2004py}). However, even though
difference between fermions and bosons is large, the total
contribution of HDM  in the central part of a halo is
only a minute fraction of the total matter density. The ratio
$\rho_{\rm HDM}/\rho_{\rm CDM}$ decreases with decreasing $r$ and
therefore HDM is most easily probed on scales
comparable to, or larger than, the virial radius where weak
lensing is most efficient.

\section{Conclusions \label{sec:conc}}

We have studied the difference between fermionic and bosonic hot
dark matter in detail. In the linear regime of structure formation,
the most important difference is that for equal contributions to
the present energy density, the mass of a scalar species is higher
than that of a fermion by a factor $3/2$. This in turn means that
scalars have a smaller free-streaming length and therefore less
suppression of fluctuation power. This effect is most pronounced
around the free-streaming scale, where the difference can be of
order $30 \to 40 \ \%$ in terms of the present day linear
power spectrum.
At present, the difference is too small to be discerned by available
cosmological data (as also
discussed in the context of nonthermally distributed neutrinos
\cite{Cuoco:2005qr}), but may be visible in
the next generation of probes for large scale structure and
in proposed wide-field weak gravitational
lensing surveys.

In the nonlinear regime, there is a fundamental difference between
fermions and bosons arising from the fact that the phase space
distribution for fermions is subject to constraints imposed by
Liouville's theorem, while for bosons there is no formal upper
bound. This difference has a direct impact on the particles'
clustering properties.  In order to probe this phenomenon, we have
studied late-time hot dark matter infall into cold dark matter
halos, and find that, in the central part of a halo, the
difference between the fermionic and bosonic overdensities can be
more than a factor five.  While this difference appears large, it
is not likely to be observable by direct means (e.g., by strong
gravitational lensing) simply because the density of hot dark
matter is so small in the central regions of halos.

Instead, we have considered the implications of hot dark matter
for the nonlinear power spectrum, $k \gwig 0.2\ h \ {\rm
Mpc}^{-1}$. These are the scales at which future weak lensing
surveys have the most statistical power.  Using a halo model
approach, we find that the early free-streaming behaviour of the
hot dark matter species remains the dominant influence on the
shape of the resulting nonlinear power spectrum. In some cases,
notably thermal relics with $g_* \gwig 10.75$, hot dark matter
infall can contribute an additional loss of power at the percent
level at $z=0$, depending on their masses and quantum statistics.
This suggests that infall can potentially play a decisive role in
high precision cosmology.  However, because both
the amount of hot dark matter infall and
the scale at which it is manifest depend crucially on the values of
$g_{*}$ and $m_{\rm HDM}$, signatures of infall may end up
unobservable in the weak lensing convergence spectrum, even with a
Large Synoptic Survey Telescope-type lensing survey,
 unless  $g_* \lwig 30$ and $1 \lwig m_{\rm
HDM}/{\rm eV} \lwig 4$.

The bottom line is that the difference in the hot dark matter
contribution arising from quantum statistics is a very interesting
example of statistical mechanics, but that it may prove to be very
difficult to observe.  On the other hand,
imprints on the matter power spectrum from different early
free-streaming behaviours extend to both the linear and the nonlinear scales.
These could well be observable with future experiments if the
mass of the hot dark matter species is sufficiently large
(say, $m_{\rm HDM} \gwig 0.2 \ {\rm eV}$).

\appendix

\section{Hot dark matter clustering \label{clustering}}

The problem of late time HDM clustering
in the presence of a CDM halo can be
treated quantitatively with the nonrelativistic Vlasov equation
(e.g., \cite{Bertschinger:1993xt}),
\begin{equation}
\label{eq:vlasov}
\frac{\partial f_i}{\partial \tau} + \frac{\bm p}{a m_{i}} \cdot
\frac{\partial f_i}{\partial {\bm x}}- a m_{i} \nabla \phi \cdot
\frac{\partial f_i}{\partial {\bm p}}=0, \qquad i={\rm CDM},\ {\rm HDM},
\end{equation}
which tracks the evolution of the phase space densities
$f_i(\bm{x},\bm{p},\tau)$ under the influence of the
peculiar gravitational potential $\phi({\bm x},\tau)$ as functions of
comoving position ${\bm x}$, momentum ${\bm p}$, and conformal time
$\tau$.
In turn, $\phi({\bm x},\tau)$ is related to
the local density fluctuations $\delta_i({\bm x},\tau)$ via
the Poisson equation,
\begin{equation}
\label{eq:poisson}
\nabla^2 \phi = \frac{4 \pi G}{a}
\sum_i \bar{\rho}_i \ \delta_i({\bm x},\tau),
\end{equation}
where
\begin{equation}
\delta_i({\bm x},\tau) \equiv \frac{\rho_i({\bm x},\tau)}{\bar{\rho}_i}-1,
\qquad \rho_i({\bm x},\tau) = m_i
\int d^3p \ f_i({\bm x},{\bm p},\tau),
\end{equation}
and $\bar{\rho}_i$ denotes the $i$th comoving mean density.

A number of methods are available for solving equations
(\ref{eq:vlasov}) and (\ref{eq:poisson})---from numerical
simulations to semi-analytical approaches under a variety of
approximation schemes---with varying degrees of accuracy.
In this work, we use a restricted
$N$-single-body method introduced in \cite{Ringwald:2004np}. Here,
test particles sampled from the initial HDM phase space
distribution are allowed to evolve, one at a time, in an external
potential $\phi({\bm x},\tau)$ due to the CDM halo alone. Gravitational
interactions between the HDM particles themselves are explicitly
ignored, which is
well justified
considering current cosmological constraints on the
HDM mass fraction today ($\Omega_{\rm HDM}/\Omega_m \lwig 0.1$).
Compared with other solution methods, our $N$-single-body approach
is computationally inexpensive relative to
full-scale $N$-body simulations,
and is yet able to reproduce essential
nonlinear effects
not captured by simpler, linear methods (e.g.,
\cite{bib:singh&ma}).
See reference \cite{Ringwald:2004np} for more details.

\section{The halo model \label{halomodel}}

The halo model supposes that all matter in the universe is
 partitioned
into distinct units, the halos.  This assumption allows one to study
the universal matter distribution in two steps: the distribution of matter
within each halo, and the spatial distribution of the halos themselves.
In its simplest application, the halo model proposes that
the matter power spectrum $P(k,z)$ be composed of two distinct terms
(e.g., \cite{Cooray:2002di}),
\begin{equation}
P(k,z) = P^{\rm 1-halo} (k,z) + P^{\rm 2-halo}(k,z),
\end{equation}
where
\begin{eqnarray}
\label{eq:p1h&p2h}
P^{\rm 1-halo} (k,z) &=& \int d M \ n(M,z)
\left(\frac{M}{\bar{\rho}_{\rm halo}} \right)^2
|u(k|M)|^2,\nonumber \\P^{\rm 2-halo} (k,z) &=& P^{\rm lin}(k,z) \left[ \int d M
\ n(M,z)\ b(M) \left(\frac{M}{\bar{\rho}_{\rm halo}} \right) u(k|M)
\right]^2,
\end{eqnarray}
describe, respectively, the correlations between two
points drawn from the same halo (``$1$-halo'') and from two different halos
(``$2$-halo'').  The former dominates on small scales (i.e., large $k$'s), while
the latter rises to prominence on large scales (i.e., small $k$'s) and
approaches the power spectrum calculated from
linear perturbation theory $P^{\rm lin}(k,z)$ as $k \to 0$.
The average (comoving) matter density $\bar{\rho}_{\rm halo}$ counts all
matter clustered in halos.
For a basic $\Lambda$CDM set-up, $\bar{\rho}_{\rm halo}$ is
well approximated by
$\bar{\rho}_{\rm halo} \simeq \bar{\rho}_m
\equiv
\Omega_m \rho_{\rm crit} $,
where $\rho_{\rm crit}$ is the present critical density.

Three pieces of information are required to complete the model.
\begin{enumerate}
\item The mass function $n(M,z)$ specifies the
comoving number density of
halos of virial mass $M$ at redshift $z$.  Following from the
Press--Schechter theory \cite{Press:1973iz}, this is usually written
as
\begin{equation}n(M,z) \ d M =
\frac{\bar{\rho}_{\rm halo}}{M} f(\nu) \ d \nu,
\end{equation}
where $f(\nu)$ is a universal function of the peak height
$\nu = \delta_{\rm sc}^2/\sigma_{\rm lin}^2(M,z).$
Here, $\delta_{\rm sc}=1.68$ is the linear overdensity at the epoch of
spherical collapse, and
\begin{equation}
\sigma_{\rm lin}^2 (M,z)\equiv \int \frac{d^3k}{(2 \pi)^3} P^{\rm lin}(k,z) |W(k
R)|^2
\end{equation}
is the rms of the linear fluctuations filtered  with a
tophat window function $W(x) = 3/x^3 (\sin x - x \cos x)$ on a scale
of $R=(3 M/4 \pi \bar{\rho}_m)^{1/3}$.  In this way, every
halo mass $M$ is associated
with a unique value of $\nu$, with large $\nu$'s corresponding large halo masses
and so forth.
A number of variants for the function $f(\nu)$ exists in the
literature.
Here, we adopt the version proposed by Sheth and Tormen \cite{Sheth:1999mn},
\begin{equation}
\label{eq:sheth-tormen}
\nu f(\nu) = A \left( 1+ (q \nu)^{-p} \right) \sqrt{\frac{q \nu}{2 \pi}}
\exp (-q \nu/2),
\end{equation}
with $p = 0.3$ and $q=0.707$, based on fits to
$N$-body simulations.
The constant $A$ is determined by mass conservation,
i.e., $\int d M \ n(M,z) \ M =\bar{\rho}_{\rm halo}$, or, equivalently,
$\int d \nu \ f(\nu) =1$.

\item The linear bias $b(\nu)$ parameterises the clustering
strength of halos relative to the underlying dark matter, and obeys
$\int d \nu \ f(\nu) \ b(\nu)=1$ by construction.
For the Sheth--Tormen
mass function (\ref{eq:sheth-tormen}), the associated bias is
\cite{Sheth:1999mn}
\begin{equation}
b(\nu) = 1 + \frac{q \nu -1}{\delta_{\rm sc}} + \frac{2p/\delta_{\rm
sc}}{1+ (q \nu)^p}.
\end{equation}

\item The function $u(k|M)$ is the Fourier transform at wavenumber $k$
of the matter distribution
$\rho(r|M)$ within a halo of mass $M$, normalised to the halo mass.
For spherically
symmetric density profiles $\rho(r|M)$ truncated at the virial radius
$r_{\rm vir}$ (to be defined below),
this is given by
\begin{equation}
u(k|M) \equiv \frac{\tilde{\rho}(k|M)}{M} = \int^{r_{\rm vir}}_{0} d r \
4 \pi r^2\ \frac{\sin k r}{k r} \frac{\rho(r|M)}{M}.
\end{equation}
In the simplest  case where only CDM and baryons cluster,
a natural choice for $\rho(r|M)$ is the
 Navarro--Frenk--White (NFW) profile
\cite{Navarro:1995iw,bib:nfw},
\begin{equation}
\label{eq:nfw}
\rho(r|M) =
\frac{\rho_s}{(r/r_s)(1+r/r_s)^2}\,,
\end{equation}
which has proven to
provide a very good description of the density run
around virialised halos in high resolution $N$-body simulations
in a variety of CDM cosmologies.
Here, $r$ is the radial distance from the halo centre. The parameters
$r_s$ and $\rho_s$ are determined by the halo's virial mass $M$ and
a dimensionless concentration $c\equiv r_{\rm vir}/r_s$ via
\begin{eqnarray}
\rho_s &=& \frac{\delta_{\rm TH}}{3} \frac{c^3}{\ln (1+c) - c/(1+c)}, \\
r_s &=& \frac{r_{\rm vir}}{c} = \frac{1}{c} \left(\frac{3}{4 \pi \delta_{\rm
TH}} \frac{M}{\bar{\rho}_m}\right)^{1/3},
\end{eqnarray}
where $\delta_{\rm TH}$ is the virial overdensity predicted by
the spherical top-hat collapse model \cite{Bryan:1997dn},
\begin{eqnarray}
\delta_{\rm TH} \simeq \frac{18 \pi^2 + 82 y - 39 y^2}{\Omega(z)},
\nonumber \\
 y = \Omega(z)-1,
 \qquad \Omega(z) = \frac{\Omega_{m}}{\Omega_{m} + \Omega_{\Lambda} a^3}.
\end{eqnarray}
In addition, a tight correlation between $c$ and $M$ has been
observed in the analysis of
\cite{bib:bullock2001},
\begin{equation}
\label{eq:cmvir}
c(z) \simeq \frac{9}{1+z} \left(\frac{M}{M^*}\right)^{-0.13}\,,
\end{equation}
where $M^*=M(\nu=1)$ denotes the $P^{\rm lin}$-dependent
nonlinear mass.

\end{enumerate}

Several modifications to the halo model have been discussed in the literature,
to account for, e.g., the effects of baryons \cite{Zhan:2004wq,White:2004kv},
 neutrino infall \cite{Abazajian:2004zh}, etc..
To include the effects of HDM infall, we follow the prescription
 of \cite{Abazajian:2004zh}
by defining
\begin{eqnarray}
M &\equiv& M_{{\rm CDM}+b} + M_{\rm HDM}, \\
u(k|M) &\equiv& \frac{\tilde{\rho}_{\rm HDM}(k|M) + \tilde{\rho}_{{\rm
CDM}+b}(k|M)}{M_{{\rm CDM}+b} + M_{\rm HDM}},
\end{eqnarray}
where $M_i$ and $\tilde{\rho}_i(k|M)$ denote, respectively, contributions from
the $i$ type particles to the total halo
mass and Fourier transform of the density profile.
Furthermore, the quantity $\bar{\rho}_{\rm halo}$ should account also for
HDM residing in halos, i.e.,
$\bar{\rho}_{\rm halo} =\bar{\rho}_{\rm CDM} + \bar{\rho}_b +\bar{\rho}_{\rm
HDM,\rm halo} \simeq (\Omega_{\rm CDM} + \Omega_{b})  \rho_{\rm crit} +
\bar{\rho}_{{\rm HDM},{\rm halo}}$, where
\begin{equation}
\bar{\rho}_{\rm HDM, halo} = \bar{\rho}_{\rm halo} \int d \nu f(\nu)
\frac{M_{\rm HDM}}{M}.
\end{equation}
Note that, in general, $\rho_{\rm HDM, halo} \ll \Omega_{\rm HDM} \rho_{\rm
crit}$, since the HDM's significant thermal energy prevents a large fraction of
particles from
participating in the
infall.


\section*{References}

\begin{thebibliography}{99}


\bibitem{Riess:1998cb}
A.~G.~Riess {\it et al.}  [Supernova Search Team Collaboration],
Astron.\ J.\  {\bf 116} (1998) 1009 [arXiv:astro-ph/9805201].

\bibitem{Perlmutter:1998np}
  S.~Perlmutter {\it et al.}  [Supernova Cosmology Project Collaboration],
  Astrophys.\ J.\  {\bf 517} (1999) 565
  [arXiv:astro-ph/9812133].

\bibitem{Spergel:2003cb}
  D.~N.~Spergel {\it et al.}  [WMAP Collaboration],
  Astrophys.\ J.\ Suppl.\  {\bf 148}  (2003) 175
  [arXiv:astro-ph/0302209].

\bibitem{bib:sdss1}
M.~Tegmark {\it et al.}  [SDSS Collaboration],
Phys.\ Rev.\ D {\bf 69} (2004) 103501 [arXiv:astro-ph/0310723].

\bibitem{Gershtein:gg}
S.~S.~Gershtein and Y.~B.~Zeldovich,
JETP Lett.\  {\bf 4} (1966) 120 [Pisma Zh.\ Eksp.\ Teor.\ Fiz.\
{\bf 4} (1966) 174].

\bibitem{Cowsik:gh}
R.~Cowsik and J.~McClelland,
Phys.\ Rev.\ Lett.\  {\bf 29} (1972) 669.




\bibitem{kolb}
E.~W.~Kolb and M.~S.~Turner, {\it The Early Universe},
Addison--Wesley (1990).
\bibitem{lesgourgues}
  M.~Viel, J.~Lesgourgues, M.~G.~Haehnelt, S.~Matarrese and A.~Riotto,
  Phys.\ Rev.\ D {\bf 71}  (2005) 063534
  [arXiv:astro-ph/0501562].

\bibitem{Hu:1997mj}
  W.~Hu, D.~J.~Eisenstein and M.~Tegmark,
  Phys.\ Rev.\ Lett.\  {\bf 80} (1998) 5255
  [arXiv:astro-ph/9712057].


\bibitem{bib:hannestad2003}
S.~Hannestad,
JCAP {\bf 0305} (2003) 004 [arXiv:astro-ph/0303076].

\bibitem{Elgaroy:2004rc}
  O.~Elgaroy and O.~Lahav,
  New J.\ Phys.\  {\bf 7} (2005) 61
  [arXiv:hep-ph/0412075].

\bibitem{Barger:2003vs}
  V.~Barger, D.~Marfatia and A.~Tregre,
  Phys.\ Lett.\ B {\bf 595} (2004) 55
  [arXiv:hep-ph/0312065].

\bibitem{Crotty:2004gm}
  P.~Crotty, J.~Lesgourgues and S.~Pastor,
  Phys.\ Rev.\ D {\bf 69} (2004) 123007
  [arXiv:hep-ph/0402049].

\bibitem{Seljak:2004xh}
  U.~Seljak {\it et al.},
  arXiv:astro-ph/0407372.

\bibitem{Fogli:2004as}
  G.~L.~Fogli, E.~Lisi, A.~Marrone, A.~Melchiorri, A.~Palazzo, P.~Serra and 
J.~Silk,  
  Phys.\ Rev.\ D {\bf 70} (2004) 113003
  [arXiv:hep-ph/0408045].

\bibitem{Tegmark:2005cy}
  M.~Tegmark,
  arXiv:hep-ph/0503257.

\bibitem{Hannestad:2005gj}
  S.~Hannestad,
  arXiv:astro-ph/0505551.



\bibitem{Hannestad:2003ye}
  S.~Hannestad and G.~Raffelt,
  JCAP {\bf 0404} (2004) 008
  [arXiv:hep-ph/0312154].

\bibitem{Moroi:1998qs}
  T.~Moroi and H.~Murayama,
  Phys.\ Lett.\ B {\bf 440} (1998) 69
  [arXiv:hep-ph/9804291].

\bibitem{Hannestad:2005df}
  S.~Hannestad, A.~Mirizzi and G.~Raffelt,
  arXiv:hep-ph/0504059.

\bibitem{Ibe:2005xc}
  M.~Ibe, K.~Tobe and T.~Yanagida,
  Phys.\ Lett.\ B {\bf 615} (2005) 120
  [arXiv:hep-ph/0503098].


\bibitem{Brandenburg:2004du}
  A.~Brandenburg and F.~D.~Steffen,
  JCAP {\bf 0408} (2004) 008
  [arXiv:hep-ph/0405158].

\bibitem{Dolgov:2005qi}
  A.~D.~Dolgov and A.~Y.~Smirnov,
  arXiv:hep-ph/0501066.

\bibitem{Dolgov:2005mi}
  A.~D.~Dolgov, S.~H.~Hansen and A.~Y.~Smirnov,
  arXiv:astro-ph/0503612.


\bibitem{Navarro:1995iw}
J.~F.~Navarro, C.~S.~Frenk and S.~D.~M.~White,
Astrophys.\ J.\  {\bf 462}  (1996) 563
[arXiv:astro-ph/9508025];

\bibitem{bib:nfw}
J.~F.~Navarro, C.~S.~Frenk and S.~D.~M.~White,
Astrophys.\ J.\
{\bf 490} (1997) 493.




\bibitem{Ringwald:2004np}
  A.~Ringwald and Y.~Y.~Y.~Wong,
  JCAP {\bf 0412} (2004) 005
  [arXiv:hep-ph/0408241].



\bibitem{tremaine1}
J.~Binney and S.~Tremaine, {\it Galactic Dynamics}, Princeton University
Press (1987).


\bibitem{Tremaine:1979we}
  S.~Tremaine and J.~E.~Gunn,
  Phys.\ Rev.\ Lett.\  {\bf 42} (1979) 407.


\bibitem{Kull:1996nx}
  A.~Kull, R.~A.~Treumann and H.~B\"{o}hringer,
  Astrophys.\ J.\  {\bf 466} (1996) L1
  [arXiv:astro-ph/9606057].


\bibitem{Madsen:1990pe}
  J.~Madsen,
  Phys.\ Rev.\ Lett.\  {\bf 64} (1990) 2744.

\bibitem{Madsen:1991mz}
  J.~Madsen,
  Phys.\ Rev.\ D {\bf 44} (1991) 999.



\bibitem{Seljak:2000gq}
  U.~Seljak,
  Mon.\ Not.\ Roy.\ Astron.\ Soc.\  {\bf 318} (2000) 203
  [arXiv:astro-ph/0001493].

\bibitem{Peacock:2000qk}
  J.~A.~Peacock and R.~E.~Smith,
  Mon.\ Not.\ Roy.\ Astron.\ Soc.\  {\bf 318} (2000) 1144
  [arXiv:astro-ph/0005010].

\bibitem{Ma:2000ik}
  C.~P.~Ma and J.~N.~Fry,
    Astrophys.\ J.\ 543 (2000) 503,
  [arXiv:astro-ph/0003343].

\bibitem{Cooray:2002di}
A.~Cooray and R.~Sheth,
Phys.\ Rept.\  {\bf 372} (2002) 1
[arXiv:astro-ph/0206508].

\bibitem{Abazajian:2004zh}
  K.~Abazajian, E.~R.~Switzer, S.~Dodelson, K.~Heitmann and S.~Habib,
  Phys.\ Rev.\ D {\bf 71} (2005) 043507
  [arXiv:astro-ph/0411552].

\bibitem{Bartelmann:1999yn}
  M.~Bartelmann and P.~Schneider,
  Phys.\ Rept.\  {\bf 340} (2001) 291
  [arXiv:astro-ph/9912508].

\bibitem{Kaiser:1991qi}
  N.~Kaiser,
  Astrophys.\ J.\  {\bf 388} (1992) 272.


\bibitem{Kaiser:1996tp}
  N.~Kaiser,
  Astrophys.\ J.\  {\bf 498} (1998) 26
  [arXiv:astro-ph/9610120].

\bibitem{Massey:2003xd}
  R.~Massey {\it et al.},
  Astron.\ J.\  {\bf 127} (2004) 3089
  [arXiv:astro-ph/0304418].

\bibitem{Refregier:2003xe}
  A.~Refregier {\it et al.},
  Astron.\ J.\  {\bf 127} (2004) 3102
  [arXiv:astro-ph/0304419].


\bibitem{Zhan:2004wq}
  H.~Zhan and L.~Knox,
  Astrophys.\ J.\  {\bf 616} (2004) L75
  [arXiv:astro-ph/0409198].



\bibitem{bib:snap}
http://snap.lbl.gov



\bibitem{bib:lsst}
http://www.lsst.org





\bibitem{Maller:1999de}
  A.~H.~Maller, L.~Simard, P.~Guhathakurta, J.~Hjorth, A.~O.~Jaunsen,
R.~A.~Flores and J.~R.~Primack,  
Gravitational Lensing,''  Astrophys.\ J.\  {\bf 533} (2000) 194
  [arXiv:astro-ph/9910207].

\bibitem{Trott:2002cr}
  C.~M.~Trott and R.~L.~Webster,
  Mon.\ Not.\ Roy.\ Astron.\ Soc.\  {\bf 334} (2002) 621
  [arXiv:astro-ph/0203196].

\bibitem{Trott:2003mn}
  C.~M.~Trott and R.~L.~Webster,
  arXiv:astro-ph/0310530.

\bibitem{Mortsell:2004py}
  E.~Mortsell, H.~Dahle and S.~Hannestad,
  Astrophys.\ J.\  {\bf 619} (2005) 733
  [arXiv:astro-ph/0406343].


\bibitem{Cuoco:2005qr}
  A.~Cuoco, J.~Lesgourgues, G.~Mangano and S.~Pastor,
  Phys.\ Rev.\ D {\bf 71} (2005) 123501
  [arXiv:astro-ph/0502465].

\bibitem{Bertschinger:1993xt}
  E.~Bertschinger, in
  {\it Les Houches Cosmology 1993}, pp.\ 273-348,
  arXiv:astro-ph/9503125.




\bibitem{bib:singh&ma}
S.~Singh and C.~P.~Ma,
Phys.\ Rev.\ D {\bf 67} (2003) 023506
[arXiv:astro-ph/0208419].


\bibitem{Press:1973iz}
  W.~H.~Press and P.~Schechter,
  Astrophys.\ J.\  {\bf 187} (1974) 425.

\bibitem{Sheth:1999mn}
  R.~K.~Sheth and G.~Tormen,
  Mon.\ Not.\ Roy.\ Astron.\ Soc.\  {\bf 308} (1999) 119
  [arXiv:astro-ph/9901122].




\bibitem{Bryan:1997dn}
  G.~L.~Bryan and M.~L.~Norman,
  Astrophys.\ J.\  {\bf 495} (1998) 80
  [arXiv:astro-ph/9710107].

\bibitem{bib:bullock2001}
J.~S.~Bullock {\it et al.},
Mon.\ Not.\ Roy.\ Astron.\ Soc.\  {\bf 321} (2001) 559
[arXiv:astro-ph/9908159].


\bibitem{White:2004kv}
  M.~White,
  Astropart.\ Phys.\  {\bf 22} (2004) 211
  [arXiv:astro-ph/0405593].





\end{thebibliography}
\end{document}